\documentclass[twocolumn,showpacs,preprintnumbers,superscriptaddress,prl]{revtex4}

\usepackage{graphicx}

\usepackage{dcolumn}

\usepackage{bm}

\begin{document}

\preprint{}

\title{Universal magnetic structure of the half-magnetization phase in Cr-based spinels}

\author{M. Matsuda}

\affiliation{Quantum Beam Science Directorate, Japan Atomic Energy Agency (JAEA), Tokai, Ibaraki 319-1195, Japan}

\author{K. Ohoyama}

\affiliation{Institute for Materials Research, Tohoku University, Katahira, Sendai 980-8577, 
Japan}

\author{S. Yoshii}

\affiliation{Institute for Materials Research, Tohoku University, Katahira, Sendai 980-8577, 
Japan}

\author{H. Nojiri}

\affiliation{Institute for Materials Research, Tohoku University, Katahira, Sendai 980-8577, 
Japan}

\author{P. Frings}

\affiliation{Laboratoire National des Champs Magn\'{e}tiques Intenses, UPR3228 CNRS-INSA-UJF-UPS, Grenoble \& Toulouse, France}

\author{F. Duc}

\affiliation{Laboratoire National des Champs Magn\'{e}tiques Intenses, UPR3228 CNRS-INSA-UJF-UPS, Grenoble \& Toulouse, France}

\author{B. Vignolle}

\affiliation{Laboratoire National des Champs Magn\'{e}tiques Intenses, UPR3228 CNRS-INSA-UJF-UPS, Grenoble \& Toulouse, France}

\author{G. L. J. A. Rikken}

\affiliation{Laboratoire National des Champs Magn\'{e}tiques Intenses, UPR3228 CNRS-INSA-UJF-UPS, Grenoble \& Toulouse, France}

\author{L.-P. Regnault}

\affiliation{CEA-Grenoble, INAC-SPSMS-MDN, 17 rue des Martyrs, 38054 Grenoble Cedex 9, France}

\author{S.-H. Lee}

\affiliation{Department of Physics, University of Virginia, Charlottesville, VA 22904-4714, USA}

\author{H. Ueda}

\affiliation{The Institute for Solid State Physics, The University of Tokyo, Kashiwa, Chiba 277-8581, Japan}

\author{Y. Ueda}

\affiliation{The Institute for Solid State Physics, The University of Tokyo, Kashiwa, Chiba 277-8581, Japan}

\date{\today}

\begin{abstract}

Using an elastic neutron scattering technique under a pulsed magnetic field up to 30 T, we determined the magnetic structure in the half-magnetization plateau phase in the spinel CdCr$_2$O$_4$. The magnetic structure has a cubic $P$4$_3$32 symmetry, which is the same as that observed in HgCr$_2$O$_4$. This suggests that there is a universal field induced spin-lattice coupling mechanism at work in the Cr-based spinels.
\end{abstract}


\maketitle

When an external magnetic field is applied on strongly interacting spin systems, novel collective phenomena may arise.\cite{Sachdev08} A well-known example is the field-induced condensation of magnons in quantum magnets.\cite{Nikuni00} Another is the field-induced fractional magnetization plateau observed in frustrated magnets.\cite{ueda2} Interest on the latter system stems from the degenerate ground state due to the triangular motif of the magnetic lattice.\cite{gingras, moessner06} For instance, for a tetrahedron made of four isotropic classical spins, any spin configuration with total zero spin can be the ground state. There is an infinite number of such configurations that satisfy the criterion. When such tetrahedra are arranged in a corner-sharing network, sometimes called the pyrochlore lattice, the ground state degeneracy becomes macroscopic, and exotic magnetic properties are expected at low temperatures.\cite{moessner98, canals98, shlnature} If the spin degree of freedom is coupled with the lattice or orbital degree of freedom, the system can undergo a phase transition at low temperature into a crystallographically distorted and magnetic ordered state.\cite{shl2000, chung2005, yamashita, Oleg02, Khomskii05, ueda97, znv2o4, Tsunetsugu, Oleg04} When an external magnetic field $(H)$ is applied to the ordered state, the fractional magnetization phase appears, and is associated with each tetrahedron having majority (minority) spins aligned parallel (antiparallel) to $H$. There can be many ways of organizing the majority and minority spins over the entire lattice, and how a certain structure can be stabilized over a wide range of $H$ and whether or not an universal ground state for each type of frustrated lattice and Hamiltonians exists are of issue.

The Cr-based spinel $A$Cr$_2$O$_4$ ($A$ = Hg, Cd, and Zn) is an ideal model system with the simplest and most frustrating spin Hamiltonian. It has a cubic $Fd\bar{3}m$ crystal structure at room temperature where the magnetic Cr$^{3+} (t_{2g}^3; s = 3/2)$ ions without orbital degeneracy form the pyrochlore lattice, and due to the edge-sharing network of CrO$_6$ octahedra, the nearest neighbor exchange interactions are dominant. Experimentally, it has been shown that despite the strong magnetic interactions evidenced by their large Curie-Weiss temperatures, $\Theta_{CW}$ = $-$32 K (Hg), $-$88 K (Cd), and $-$390 K (Zn), the system remains in a spin-liquid state down to temperatures $(T)$ much lower than $|\Theta_{CW}|$, indicating the presence of strong frustration.\cite{ueda2} Upon further cooling, it undergoes a transition at 6 K (Hg), 7.8 K (Cd), and 12.5 (Zn) due to spin-lattice coupling into a magnetically long range ordered state.~\cite{shl2000,chung2005,matsuda0} The nature of the magnetic structure and lattice distortion is different for different $A$ site ions: the symmetry of the low $T$ crystal structure is orthorhombic $Fddd$, tetragonal $I4_1/amd$, and tetragonal $I\bar{4}m2$ for Hg, Cd and Zn, respectively. Their magnetic structures also have different characteristic wave vectors: two commensurate wave vectors $\mathbf{Q}\rm_m =$ (1,0,1/2), (1,0,0) for Hg \cite{matsuda0}, a single incommensurate $\mathbf{Q}\rm_m = (0,\delta,1)$ for Cd \cite{chung2005}, and two commensurate $\mathbf{Q}\rm_m =$ (1/2,1/2,0), (1,0,1/2) for Zn \cite{ji2009}. These indicate that the Cr spinels are very close to a critical point surrounded by several different spin structures in phase space, and the microscopic mechanism of the zero-field spin-lattice coupling depends on the delicate balancing acts between spin and lattice degrees of freedom that vary with the $A$ site ion. 

When an external field, $H$, is applied, the Cr spinel undergoes a phase transition into a half-magnetization plateau phase at $H_{c1} = 10$ T for Hg, 28 T for Cd, and 120 T for Zn~\cite{ueda2,ueda,kojima}, suggesting that each tetrahedron has three up (majority) and one down (minority) spins (3:1 constraint). The field-induced magnetic and chemical structure of HgCr$_2$O$_4$ were determined to have the $P$4$_3$32 symmetry or its mirror 
image $P4_132$ since its $H_{c1}$ is within the steady magnet capability available at neutron facilities.\cite{matsuda0} A question that arises is whether or not the nature of the field-induced phase in the Cr spinels varies with different $A$ ions as it does for the zero-field spin-lattice coupling. Studying other Cr spinels has however been impossible because of their high critical fields that are beyond the current steady magnet technology available at neutron facilities. Very recently, a pulsed magnet capability that can go up to 30 T has been implemented at neutron facilities, opening up a new research opportunity in this field. 

Here, we report our elastic neutron scattering measurements on CdCr$_2$O$_4$ under the pulsed magnetic field to study its half-magnetization plateau phase. We show that for $H > H_{c1} = 28$ T the incommensurate magnetic peaks disappear while new peaks appear with the characteristic wave vector of $\mathbf{Q}\rm_m =$(1,0,0) but not at the (2,$\bar{2}$,0) point. This clearly demonstrates that the half-magnetization plateau phase of CdCr$_2$O$_4$ has the same $P4_332$ magnetic structure as that of HgCr$_2$O$_4$. Our results suggest that the observed $P4_332$ state might be the generic ground state of the field-induced phase of the Cr-spinels, despite their different crystal and magnetic structures observed at $H = 0$.

\begin{figure}
\includegraphics[width=8.5cm]{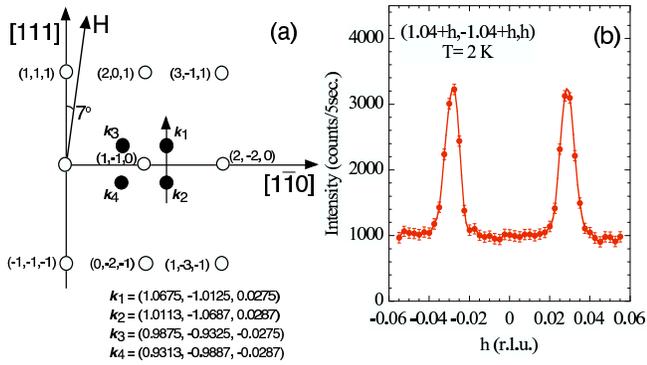}
\caption{(Color online) (a) A schematic diagram of the [111] and [1\={1}0] horizontal scattering plane that was used for our neutron scattering measurements. The external pulsed magnetic field, $H$, was applied horizontally 7$^\circ$ away from the [111] direction as shown by an arrow. The open circles represent commensurate wave vector positions, and the filled circles represent the (0,$\delta$,1) incommensurate magnetic Bragg positions observed for zero magnetic field. (b) Our elastic neutron scattering data taken with $H = 0$ at 2 K along the (1.04+$h$,$-$1.04+$h$,$h$) as shown by an arrow in (a).}
\label{diagram}
\end{figure}

A single crystal that has a shape of a thin plate ($\sim$4 mm$\times$4 mm$\times$0.2 mm) and weighs $\sim$40 mg was used. Since natural Cd has a large neutron absorption cross section, a single crystal enriched with $^{114}$Cd isotope was used for our measurements. The elastic neutron scattering experiments were carried out on the thermal neutron triple-axis spectrometers IN22 at Institut Laue-Langevin (ILL). The incident and final neutron energies were fixed to $E\rm_i$=34.8 meV. Contamination from higher-order neutrons was effectively eliminated by a PG filter. The 40 mg single crystal was mounted with the [111] and [1\={1}0] axes in the horizontal scattering plane. A small magnet coil made of CuAg wire was mounted surrounding the crystal on a cryostat insert and the apparatus was cooled in a standard $^4$He cryostat.~\cite{ohoyama1,ohoyama2,yoshii} The magnet coil in the cryostat was connected to a transportable capacitor bank that resided outside the cryostat. A half-sine shaped pulsed magnetic field of 8 msec duration was generated by using a capacitor bank.\cite{frings} The magnetic field was measured by a set of pick-up coils installed around the sample. The magnet coil limited the accessible scattering angle below 30$^\circ$, which allowed us to reach only two commensurate reflections at (1, $-$1,0) and (2, $-$2,0), and incommensurate peaks around (1, $-$1,0). The pulsed measurements were performed more than 100 times at each reflection to obtain reasonable statistics.

\begin{figure}
\includegraphics[width=8.5cm]{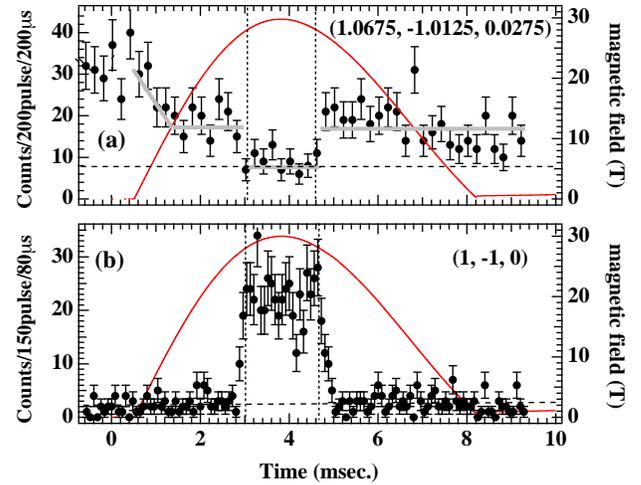}
\caption{(Color online) Time dependence of the magnetic field (solid red lines) and neutron counts (filled circles) measured at (1.0675, $-$1.0125, 0.0275) and (1, $-$1, 0) reflections at the initial temperature $T$=2.5 K. The corresponding magnetic field is shown on the right y-axis. The measurements were performed over 200 and 150 magnetic field pulses for (1.0675, $-$1.0125, 0.0275) and (1, $-$1, 0) reflections, respectively, and the data were summed. Binning times were 200 and 80 $\mu$s for (1.0675, $-$1.0125, 0.0275) and (1, $-$1, 0) reflections, respectively. The vertical dotted lines are drawn at the times corresponding nominally to the critical field, $H_c = 28$ T. The horizontal dashed lines represent the background levels determined at wave vectors away from the reflection positions. The gray thick lines are guide for eyes.}
\label{time_dependence}
\end{figure}

At zero magnetic field, CdCr$_2$O$_4$ shows a spiral magnetic order with a single characteristic wave vector of $\mathbf{Q}\rm_m$ = (0, $\delta$, 1) or ($\delta$,0,1) where $\delta\sim 0.09$, accompanied by a tetragonal distortion below $T\rm_N = 7.8$ K. \cite{chung2005,matsuda1} Thus, the incommensurate (IC) wave vectors within the accessible scattering angle of 30$^\circ$ are (1,$-1\pm\delta$,0) and (1$\pm\delta,-1$,0). Even though these peaks are out of the scattering plane separated by $\sim$ 0.025 \AA$^{-1}$, they could be detected in our measurements because the full-width-of-the-half-maximum of the vertical instrumental resolution was 0.144 \AA$^{-1}$. We performed elastic scans around the (1, $-$1,0) point, and found superlattice peaks at four positions shown in Fig. 1 (a). In Fig. 1 (b), a typical elastic scan taken along the (1,1,1) direction centered at (1.04, $-$1.04,0) show two peaks at ${\bf k}_1$ and ${\bf k}_2$ of Fig. 1 (a). The four peak positions, ${\bf k}_1$, ${\bf k}_2$, ${\bf k}_3$, and ${\bf k}_4$, correspond to the IC peaks at $(1.09, $-$1,0), (1, $-$1.09,0), (1, $-$0.91,0),$ and $(0.91, $-$1,0)$ projected on the scattering plane.

\begin{figure}
\includegraphics[width=9cm]{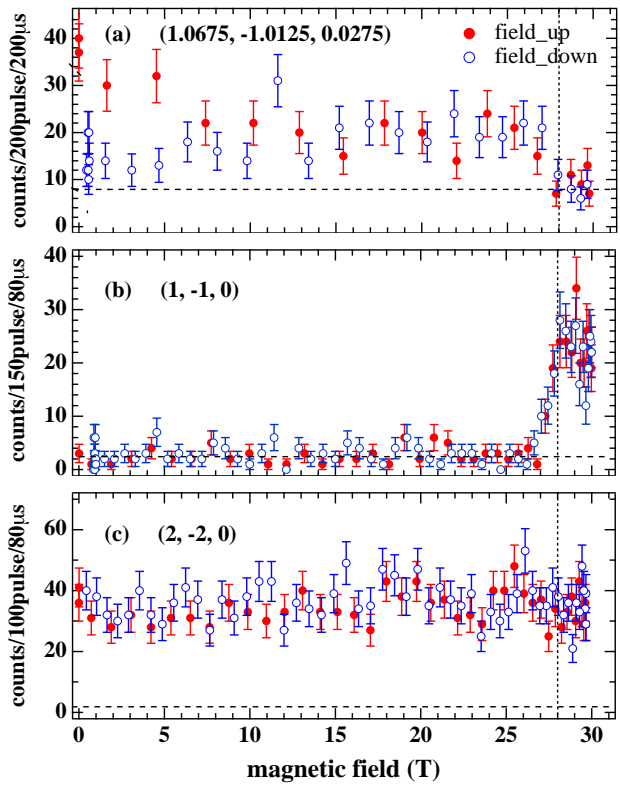}
\caption{(Color online) Magnetic field dependence of the peak intensity of the (a) (1.0675, $-$1.0125, 0.0275), (b) (1, $-$1, 0), and (c) (2, $-$2, 0) reflections measured at $T$=2.5 K with the ascending (filled circles) and descending (open circles) field. For better statistics, binning times were 200, 80, and 80 $\mu$s for (1.0675, $-$1.0125, 0.0275), (1, $-$1, 0), and (2, $-$2, 0) reflections, respectively. The vertical dotted lines are drawn at the times corresponding nominally to the critical field, $H_c = 28$ T. The horizontal dashed lines represent background intensity.}
\label{field_dependence}
\end{figure}

Figure \ref{time_dependence}(a) shows the time dependence of the elastic neutron scattering intensity measured at the IC magnetic peak of (1.0675, $-$1.0125, 0.0275) at 2.5 K. The peak intensity initially decreases gradually up to 1 ms that corresponds to $H = 8$ T, and remains constant at 20 counts per 200 pulses up to 3 ms, after which the intensity sharply decreases to background level. The 3 ms time corresponds to $H =$ 28 T, which is consistent with the critical field observed in the previous bulk magnetization measurements\cite{ueda2}. The magnetic field reaches the maximum 29.6 T at 3.9 ms after which $H$ decreases. The IC magnetic signal remains zero up to 4.6 ms ($H = 28$ T) after which the intensity increases back to the intermediate level at $H = 8$ T but not to the original intensity at $H = 0$ T. The hysteretic behavior originates from the magnetic domain orientation and is consistent with the previous result.~\cite{matsuda1} While waiting about 8 minutes for the next current pulse, the sample was warmed up to 20 K $> T_N$ and cooled back down to 2.5 K to recover the original zero-field intensity.

In order to find out where the elastic magnetic intensity that disappeared at the IC wave vector for fields $H > 28$ T transferred to, we performed similar measurements at a commensurate $\bf{Q}$=$(1,-1,0)$ position. Figure \ref{time_dependence}(b) and Figure \ref{field_dependence}(b) show the results as a function of time and field. When a magnetic field was injected, no signal was initially observed at (1, $-$1,0) for 3 ms at which point the intensity suddenly increased due to the first-order nature of the field-induced phase transition. The commensurate magnetic intensity remained non-zero over exactly the same range of time (and field) over which the IC magnetic signal went down to zero. Our results indicate that as CdCr$_2$O$_4$ enters the half-magnetization plateau state upon application of an external magnetic field, the magnetic structure changes from the incommensurate spiral to a commensurate collinear spin structure with $\bf{Q_m}$=(1,0,0). Once the 3:1 constraint is imposed, the (100)-type reflections are consistent with two non-equivalent spin configurations for the pyrochlore lattice: one with cubic $P4_332$ symmetry (Fig. 4 (a)) and the other with rhombohedral $R\bar{3}m$ symmetry (Fig. 4 (b)).\cite{matsuda0} To distinguish between the two models, we also performed similar pulsed field measurements at $(2,\bar{2},0)$ at which point the $R\bar{3}m$ structure should produce magnetic Bragg scattering while the $P4_332$ structure would not. At $H$=0, nuclear Bragg intensity is observed at $(2,\bar{2},0)$. As shown in Fig. 4(c), the $(2,\bar{2},0)$ intensity does not change as the system enters the half-magnetization phase. Thus, we conclude that the half-magnetization spin state of CdCr$_2$O$_4$ has the same $P4_332$ spin structure as observed in HgCr$_2$O$_4$. This is rather surprising because the two systems have quite different ground states at $H = 0$.

\begin{figure}
\includegraphics[width=8.5cm]{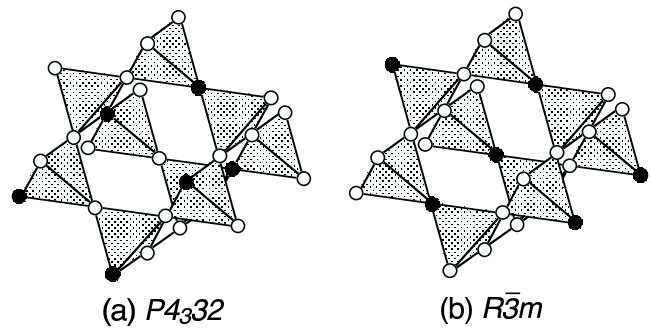}
\caption{Magnetic structures with cubic $P$4$_3$32 (a) and rhombohedral $R\bar{3}m$ (b) symmetries. Open and filled circles represent up and down spins, respectively. Each tetrahedron has three up and one down spins.}
\label{structure}
\end{figure}

Why is the $P4_332$ spin structure favored as the ground state of the half-magnetization phase in Cr-spinels, regardless of their different zero field ground states? There are many ways of arranging the pyrochlore lattice with tetrahedra holding the 3:1 constraint because there is considerable freedom in choosing the location of the down spin on each tetrahedron. The most relevant Hamiltonian for the Cr-spinels is the nearest neighbor exchange interaction that is sensitive to the bond distance minus an elastic energy associated with the displacements of the magnetic atoms. This effective hamiltonian has already been investigated theoretically as an Einstein phonon model, showed that maximizing the displacements or minimizing the Hamiltonian occurs in a unique bending pattern of tetrahedra that has the $P4_332$ symmetry.\cite{bergman2} Previous combined neutron and synchrotron x-ray measurements on HgCr$_2$O$_4$ showed that at the field-induced phase transition the crystal structure indeed becomes cubic with the $P4_332$ symmetry.\cite{matsuda0} A recent synchrotron x-ray diffraction experiments on CdCr$_2$O$_4$ under pulsed magnetic field showed that the crystal structure of the half-magnetization phase is cubic as well.~\cite{inami} These results suggest that the simple effective Hamiltonian describes the physics of the field-induced phase transition into the half-magnetization plateau phase in the Cr-spinels.

In summary, our elastic neutron scattering experiments on a single crystal of CdCr$_2$O$_4$ under a pulsed magnetic field up to 30 T showed that the Cr-spinels, $A$Cr$_2$O$_4$ with nonmagnetic $A$ ions have a unique field-induced half-magnetization state, regardless of their different zero-field ground states. This is consistent with the theoretical prediction based on the simplest Hamiltonian for the spin-lattice coupling with the nearest neighbor exchange interaction and an elastic energy term.\cite{bergman2} In addition, our study demonstrates that with the new pulsed magnet capability up to 30 T now available at neutron facilities, new research opportunities in the field of frustrated magnetism open up.

This work was partially supported by Grant-in-Aid for Scientific Research on priority Areas "High Field Spin Science in 100T" (No.451), "Novel States of Matter Induced by Frustration" (19052004 and 19052008) from the Ministry of Education, Culture, Sports, Science and Technology (MEXT) of Japan, and by ICC-IMR center of Tohoku University. S.H.L. is supported by U.S. DOE through DE-FG02-07ER46384.

\end{document}